\documentclass[15pt]{article}

\usepackage{arxiv}

\usepackage[utf8]{inputenc} 
\usepackage[T1]{fontenc}    
\usepackage{hyperref}       
\usepackage{url}            
\usepackage{booktabs}       
\usepackage{amsfonts}       
\usepackage{nicefrac}       
\usepackage{microtype}      
\usepackage{lipsum}
\usepackage{amsmath}
\usepackage[labelfont=bf]{caption}
\usepackage{siunitx}
\usepackage{bm}
\usepackage{color, colortbl} 
\usepackage{multirow}
\usepackage{graphicx}
\usepackage{setspace}
\usepackage[superscript]{cite}

\begin{document}
\captionsetup[figure]{labelfont={bf},name={Fig.},labelsep=period}
\vspace*{0.35in}


\begin{flushleft}
{\Large
\textbf\newline{\bf{Isotopic study of Raman active phonon modes in \bm{$\beta-\mathrm{Ga}_2\mathrm{O}_{3}$}}}
}
\newline
\\
B. M. Janzen\textsuperscript{1},
P. Mazzolini \textsuperscript{2},
R. Gillen\textsuperscript{3},
A. Falkenstein\textsuperscript{4},
M. Martin\textsuperscript{4},
H. Tornatzky\textsuperscript{1},
O. Bierwagen\textsuperscript{2},
M. R. Wagner\textsuperscript{1,*}
\line(1,0){470}
\\
\vspace{.3cm}
\bf{1:} \normalfont Technische Universität Berlin, Institute of Solid State Physics, Hardenbergstraße 36, 10623 Berlin, Germany
\\
\bf{2:} \normalfont Paul-Drude-Institut für Festkörperelektronik, Leibniz-Institut im Forschungsverbund Berlin e.V, Hausvogteiplatz 5-7, 10117 Berlin, Germany
\\
\bf{3:} \normalfont Institute of Condensed Matter Physics, University of Erlangen-Nürnberg, Staudtstraße 1, 91058 Erlangen, Germany
\\
\bf{4:} \normalfont Institute of Physical Chemistry, RWTH Aachen University, Landoltweg 2, 52074 Aachen, Germany
\\
\vspace{.5cm}
* markus.wagner@physik.tu-berlin.de
\line(1,0){470}

\end{flushleft}

\begin{abstract}
Holding promising applications in power electronics, the wide band gap material gallium oxide has emerged as a vital alternative to materials like GaN and SiC. 
The detailed study of phonon modes in $\beta$-Ga$_{2}$O$_{3}$ provides insights into fundamental material properties such as crystal structure and orientation and can contribute to the identification of dopants and point defects.
We investigate the Raman active phonon modes of $\beta$-Ga$_{2}$O$_{3}$ in two different oxygen isotope compositions ($^{16}$O,$^{18}$O) by experiment and theory: By carrying out polarized micro-Raman spectroscopy measurements on the (010) and ($\bar{2}$01) planes, we determine the frequencies of all 15 Raman active phonons for both isotopologues. The measured frequencies are compared with the results of density functional perturbation theory (DFPT) calculations. In both cases, we observe a shift of Raman frequencies towards lower energies upon substitution of $^{16}$O with $^{18}$O. By quantifying  the  relative frequency shifts of the individual Raman modes, we identify the atomistic origin of all modes (Ga-Ga, Ga-O or O-O) and present the first experimental confirmation of the theoretically calculated energy contributions of O lattice sites to Raman modes. We find that oxygen substitution on the O$_{\mathrm{II}}$ site leads to an elevated relative frequency shift compared to O$_{\mathrm{I}}$ and O$_{\mathrm{III}}$ sites. This study presents a blueprint for the future identification of different point defects in Ga$_{2}$O$_{3}$ by Raman spectroscopy.
\end{abstract}

\section{Introduction}

Gallium oxide (Ga$_{2}$O$_{3}$) is a semiconductor material with a wide bandgap of about 4.8 eV, \cite{Ueda1997,Yamaga} enabling applications in power-electronics devices such as rectifiers,\cite{Pearton,Pearton2018} metal-semiconductor field-effect transistors (MESFETs),\cite{Higashiwaki,Higashiwaki2} metal oxide semiconductor field-effect transistor (MOSFETs)\cite{Higashiwaki2017,Pearton,Pearton2018} or deep-UV photo detectors\cite{Kokubuna,Oshima}. The material may, depending upon temperature and pressure, exist in five different polymorphs: $\alpha,\beta,\gamma,\kappa$ (mostly referred as $\epsilon$)\cite{Cora} and $\delta$.\cite{Roy,Yoshioka,Weckenstern,Stepanov,He_1,He_2,Kroll,Playford,Furthmueller,Pearton} 
The majority of the Ga$_{2}$O$_{3}$ related research focuses on the $\beta$-phase as it is the thermodynamically most stable phase\cite{Roy,Yoshioka,Playford} and can be produced as bulk single crystals by different growth techniques such as the Czochralski method,\cite{Galazka2010,Galazka2014,Irmscher2011,Vasyltsiv} floating-zone,\cite{Ueda1997,Villora2002,Suzuki2007,Ohira2008,Zhang2006} edge-defined film fed (EFG),\cite{Kuramata2016,Vasyltsiv} or Bridgman (horizontal or vertical, HB and VB)\cite{Nikolaev2017,Mohamed2012,Suzuki2004} growth methods. 

\subsection{Crystal Structure and Electronic Properties}

\begin{figure}[h]
\centering
\includegraphics[width=90mm]{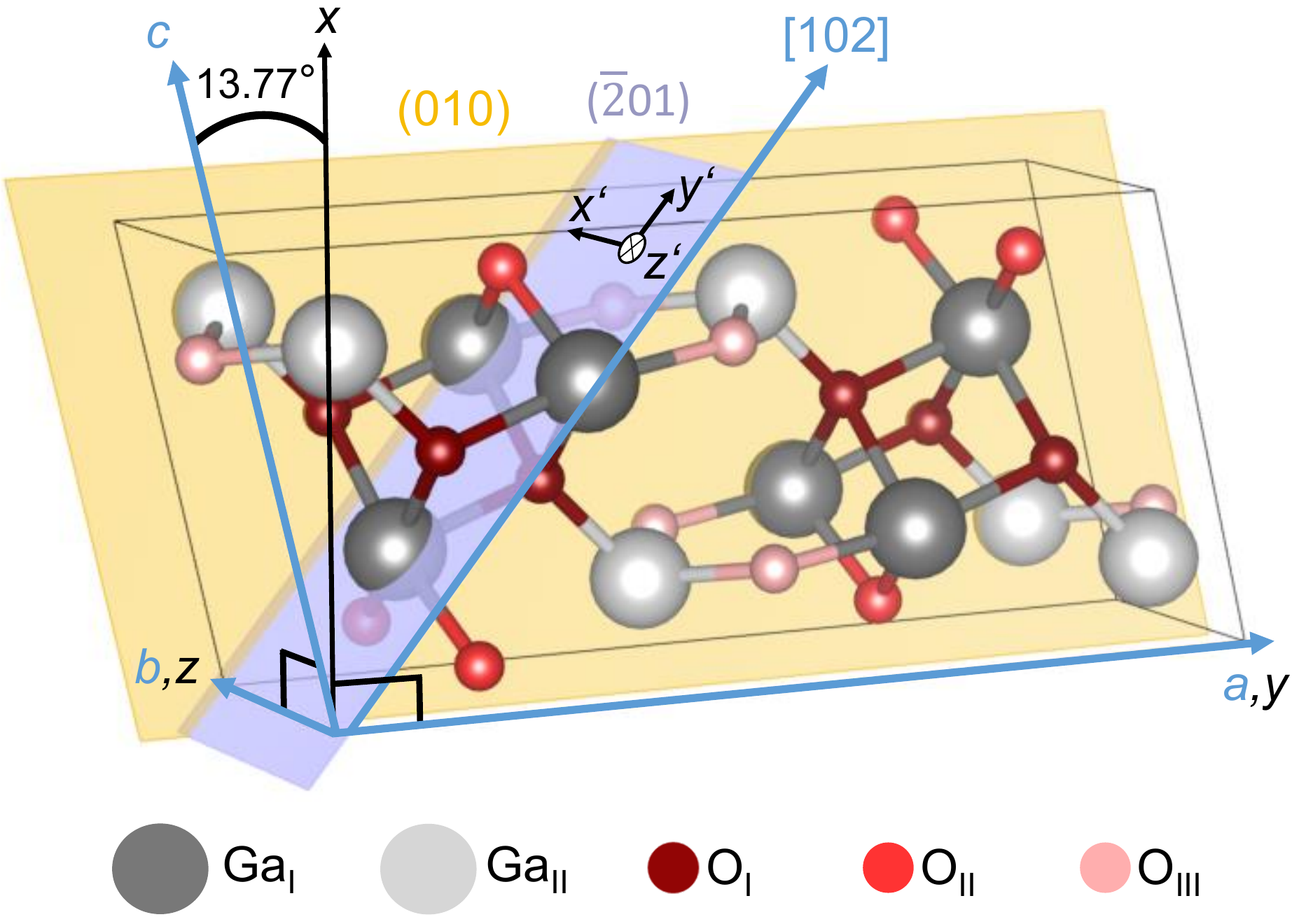}
\caption{Unit cell of monoclinic $\beta$-Ga$_{2}$O$_{3}$. Two types of Ga (Ga$_{\mathrm{I}}$ and Ga$_{\mathrm{II}}$) and three types of O (O$_{\mathrm{I}}$, O$_{\mathrm{II}}$ and O$_{\mathrm{III}}$) lattice sites are illustrated. By aligning the $y$ and $z$ axes of a Cartesian coordinate system (black) along the crystallographic $a$ and $b$ axes (blue), the $x$ and $c$ axes confine an angle of 13.77$^{\circ}$. The (010) plane (yellow) is spanned by the $a$ and $c$ axes. Cartesian coordinates $x'\parallel b$ and $y'\parallel [102]$ parameterize the ($\bar{2}01$) plane (violet), with $z'$ denoting the surface normal. }\label{Figure: Unit Cell of beta gallium oxide}
\end{figure}

The crystal structure of the $\beta$-polymorph (Fig. \ref{Figure: Unit Cell of beta gallium oxide}) is monoclinic (space group: $C_{2\mathrm{h}}^{3}$; $C2/m$)\cite{Furthmueller,Yoshioka,Geller} with lattice parameters \mbox{$a=12.29$ \si{\angstrom},} $b=3.05$ \si{\angstrom}, \mbox{$c=5.81$ \si{\angstrom}} and the monoclinic angle of 103.77$^{\circ}$ between the crystallographic $a$ and $c$ axes.\cite{Furthmueller} 

The commercial availability of $\beta$-Ga$_{2}$O$_{3}$ substrates with different surface planes enables lattice-matched strain-free homoepitaxial growth of $\beta$-Ga$_{2}$O$_{3}$ films with low defect densities.\cite{Mazzolini,Sasaki2012,Higashiwaki2016,Rafique2018,Tsai2010,Wagner2014,Murakami2014,Lee2016}
Nonetheless, even in the case of homoepitaxy, different substrate orientations can result in the formation of structural defects in the deposited layers, e.g. twin domains in (100) and ($\bar{2}$01) films.\cite{Wagner2014,Mazzolini} 
In order to overcome this problem, homoepitaxial growth of $\beta$-Ga$_{2}$O$_{3}$ is mostly performed on (010) oriented substrates,\cite{Sasaki2012} although viable alternatives such as (001) homoepitaxy\cite{Mazzolini} or the employment of proper offcuts in (100) oriented substrates\cite{Schewski2019} were recently reported.
\newline 
Previous works have demonstrated the possibility of extrinsic n-doping by substituting Ga atoms with electron donors like Si,\cite{Villora} Sn,\cite{Orita} Ge\cite{Ahmadi} and Nb\cite{Zhou}. To date, the reported charge carrier densities in $\beta$-Ga$_{2}$O$_{3}$ bulk crystals\cite{Siah,Onuma_2} and thin films\cite{Ahmadi,Zhou,Baldini} have rarely exceeded high \mbox{$10^{19}$cm$^{-3}$}. Free carrier concentrations are commonly limited by point defects and structural defects. Theoretical works\cite{Varley,Lany,Lyons,Lee,Zacherle,Krystos,Deak,Peelaers} have addressed the effects of Ga and O point defects in gallium oxide on the electrical and optical properties. Using hybrid functional calculations, it was revealed that oxygen vacancies (VO) are deep donors with ionization energies above 1 eV and hence do not contribute to the observed n-type conductivity in unintentionally doped $\beta$-Ga$_{2}$O$_{3}$.\cite{Varley}
Succeeding studies found that all other native defects, except for gallium interstitials (Ga$_{i}$), are deep, too.\cite{Krystos,Zacherle,Deak} While Ga$_{i}$ may act as shallow donors, their high mobility, coupled with large formation energies under n-type conditions inhibit their contribution to the electrical conductivity.\cite{Peelaers}
Whereas intrinsic defects should not be the main source of electrical conductivity, different impurities including H, Si, Ge, Sn, F and Cl could act as extrinsic shallow donors.\cite{Lany,Varley} Moreover, a large concentration of point defects can also affect the mobility of free charge carriers. Nonetheless, from an experimental point of view both (i) the control over the formation of different point defects during the material synthesis and (ii) their unambiguous identification is extremely difficult to achieve, but potentially of paramount importance for fully exploiting the material potential. In order to address this challenge, we conduct a combined experimental and theoretical study of the impact of different lattice atoms on phonon modes in $\beta$-Ga$_{2}$O$_{3}$ isotopologues.

\subsection{Lattice Dynamics}

Polarized micro-Raman spectroscopy constitutes a powerful tool that enables e.g. the study of strain in epitaxial thin films,\cite{Wagner} the detection of dopants and point defects via local vibrational modes,\cite{Kaschner} the identification of lattice sites via angular resolved measurements, \cite{Kranert} and the determination of the thermal conductivity via Raman thermometry in its one and two laser implementations.\cite{Reparaz,Chavez-Angel} 
Symmetries and spectral positions of Raman active phonons of $\beta$-Ga$_{2}$O$_{3}$ in the natural $^{16}$O isotopic composition have been calculated theoretically and verified experimentally in a number of publications.\cite{Kranert,Dohy,Machon,Liu} Infrared active phonons were investigated by IR ellipsometry.\cite{Schubert} 
The influence of lattice expansion was studied by temperature dependent Raman spectroscopy of $\beta$-Ga$_{2}$O$_{3}$.\cite{Dohy} Employing the valence force field calculation, the authors indentified three categories of Raman active phonons with respect to the motions of Ga and O atoms. 
The transition of the $\beta$- to the $\alpha$-phase under high pressure was investigated by carrying out Raman spectroscopy in diamond anvil cells.\cite{Machon}
Raman tensor elements were determined in angular-resolved measurements\cite{Kranert} using a modified Raman tensor formalism proposed in a preceding publication.\cite{Kranert_Grundlagen}
\newline Raman vibrations may be excited by oscillations of Ga or O atoms. By calculating the displacements of the individual O or Ga lattice sites, density functional perturbation theory (DFPT) calculations enable the identification of the atomistic origins of vibrational modes (Ga-Ga, Ga-O or O-O). From an experimental point of view, the identification of individual lattice sites in materials with large unit cells is a challenging endeavor. A powerful tool is the usage of different isotopes and investigate their impact on the vibrational properties of a material. 
This approach was successfully applied in TiO$_{2}$ enabling the experimental identification of Raman modes without any contribution of oxygen lattice vibrations as evidenced by the unchanged frequency of the $E_{g}(1)$
and $B_{1g}(1)$ modes in anatase TiO$_{2}$.\cite{Kavan,Frank} 
\newline Alternatively, the introduction of different isotopes of the same dopant can provide an unambiguous identification of dopant related local vibrational modes. The substitution of N on O lattice sites in ZnO produced a pair of additional modes at $\nu\approx 274$ and 510 cm$^{-1}$,\cite{Kaschner2002,Friedrich2007,friedrich_later,Gluba2013} irrespective of the implanted N isotope\cite{Artus2007,Wang2006}. Moreover, since the doping of ZnO with Ga, Fe, Sb, Li and Al yielded the same vibrational modes,\cite{Bundesmann2003} their occurrence was tentatively attributed to a Zn$_{\mathrm{I}}$-N$_{\mathrm{O}}$ or Zn$_{\mathrm{I}}$-O$_{\mathrm{I}}$ complex.\cite{friedrich_later} 
By investigating undoped and nitrogen-doped ZnO thin films in different Zn isotope compositions, the presence of the 274 cm$^{-1}$ mode was eventually revealed to be related to interstitial Zn clusters depending on the surface polarity of ZnO.\cite{Gluba2013}
This experimental approach could represent an important milestone for the future identification of different point defects in Ga$_{2}$O$_{3}$ layers as a function of different synthesis (e.g. deposition, annealing) conditions.\cite{Mazzolini2}
\newline In this work, confocal and cross-section micro-Raman spectroscopy is used to investigate the phonon frequencies of $\beta-\mathrm{Ga}_{2}\mathrm{O}_{3}$ in the natural $^{16}$O and $^{18}$O isotope compositions. The acquired experimental data are complemented with the results of DFPT calculations. By quantifying the relative frequency shifts of the individual Raman modes, we discuss the origins of vibration dynamics (Ga-Ga, Ga-O or O-O) and present the first experimental confirmation of the theoretically calculated energy contribution of O sites to Raman modes.
The results of our study shall open the possibility of analyzing O-related point defects in $\beta$-Ga$_{2}$O$_{3}$ by Raman spectroscopy.

\section{Experimental and theoretical methods}

A $\beta-\mathrm{Ga}_{2}\mathrm{O}_{3}$ layer was homoepitaxially deposited (deposition time 445 min) on top of an unintentionally doped (010)-oriented substrate with In-mediated metal-exchange catalysis (MEXCAT)\cite{Vogt2017,Bierwagen2020,Mazzolini2019} in an MBE chamber equipped with an oxygen-plasma source run at a power of 300 W. For this deposition run, nominally 97.39\% isotopically enriched $^{18}$O bottle was employed to provide an oxygen flux of 0.38 standard cubic centimeter per minute (sccm) during the layer growth. For gallium a beam equivalent pressure (BEP) of \mbox{BEP$_{\mathrm{Ga}}$ = $1.27\cdot 10^{-7}$ mbar} was used (equivalent to a particle flux of \mbox{$\Phi_{\mathrm{Ga}}$ = 2.2 nm$^{-2}$ s$^{-1}$),} while the additional In-flux necessary to allow the catalytic growth of the layer at a substrate temperature of T$_{\mathrm{g}}$ = 900 $^{\circ}$C was set to \mbox{$\Phi_{\mathrm{In}}$ = 1/3 $\Phi_{\mathrm{Ga}}$} (BEP$_{\mathrm{In}}=5.2\cdot10^{-8}$ mbar). 
\newline The $^{18}$O isotope fraction, $n^*$, within the sample was determined by time-of-flight secondary ion mass spectrometry (ToF-SIMS) depth profiles on a ToF-SIMS IV machine (IONTOF GmbH, Münster, Germany). Measurements were performed with a 25 kV Ga$^+$ analysis beam and a 2 kV Cs$^+$ sputter beam. The isotope fraction is directly accessible from the $^{18}$O and $^{16}$O intensities by $n^*=I_\mathrm{^{18}O}/(I_\mathrm{^{16}O}+I_\mathrm{^{18}O})$. Crater depths were analyzed by interference microscopy with a WYKO NT1100 (Veeco Instruments Inc., USA).
\newline Raman scattering at room temperature (293 K) was excited by a 632.816 nm He-Ne laser on a LabRAM HR 800 spectrometer (Horiba Jobin-Yvon, France). The laser beam was focused onto the sample using a 100x Olympus objective with a numerical aperture (NA) of 0.9, with the scattered light being collected in backscattering geometry. Backreflected and elastically scattered light (Rayleigh component) was filtered using an ultra low frequency filter (ULF) unit and then spectrally-dispersed by a monochromator spectrometer with a grating of 1800 lines/mm. The light was detected by a charge-coupled device (CCD).
The sample was placed beneath the objective with a respective surface's normal parallel to the direction of light propagation. A $\lambda/2$ wave plate in the excitation was set at 0$^{\circ}$ or 45$^{\circ}$ to polarize the incident light parallel or crossed with respect to the scattered light, which was selected using a fixed polarizer in the detection. 
Prior to each measurement, the Raman spectrometer was calibrated using the silicon $T_{2g}$-peak at \mbox{520.7 cm$^{-1}$.}

Simulations of the lattice vibrations were performed within the frame of density functional perturbation theory (DFPT) on the level of the local density approximation (LDA) as implemented into the Quantum Espresso suite~\cite{qe}. The Ga($3s$,$3p$,$3d$) and the O($2s$,$2p$) states were treated as valence electrons using multi-projector optimized normconserving Vanderbildt (ONCV) pseudopotentials~\cite{oncvpsp} from the Pseudo Dojo repository~\cite{pseudodojo}, where we used a large cutoff of the planewave basis set of 180 Ry. All reciprocal space integrations were performed by a discrete $k$-point sampling of 7x7x7 points in the Brillouin zone. We fully optimized the atomic positions and cell parameters of the primitive cell of $\beta$-Ga$_2$O$_3$ until the 
residual forces between atoms and the cell stress were smaller than 0.0025 eV/\AA and 0.01 GPa, respectively. The threshold for the total energy was set to 10$^{-15}$ Ry, which ensured tightly converged interatomic forces for the geometry optimization and of the ground state density and wavefunctions for the DFPT calculations.   

\begin{figure}[!htb]
    \centering
    \begin{minipage}{.5\textwidth}
        \flushleft
        \includegraphics[width=81mm]{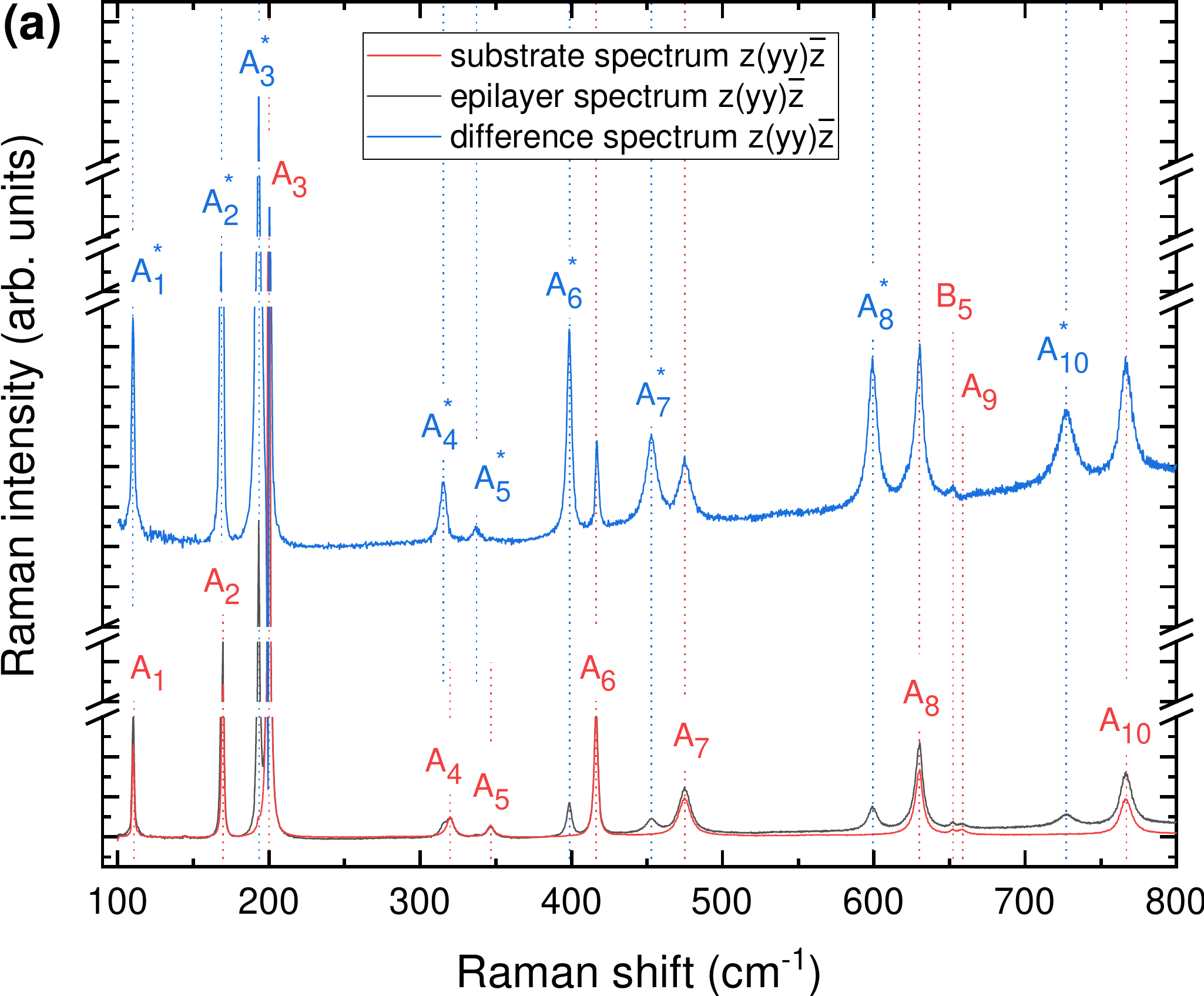}
    \end{minipage}%
    \begin{minipage}{0.5\textwidth}
        \flushright
        \includegraphics[width=81mm]{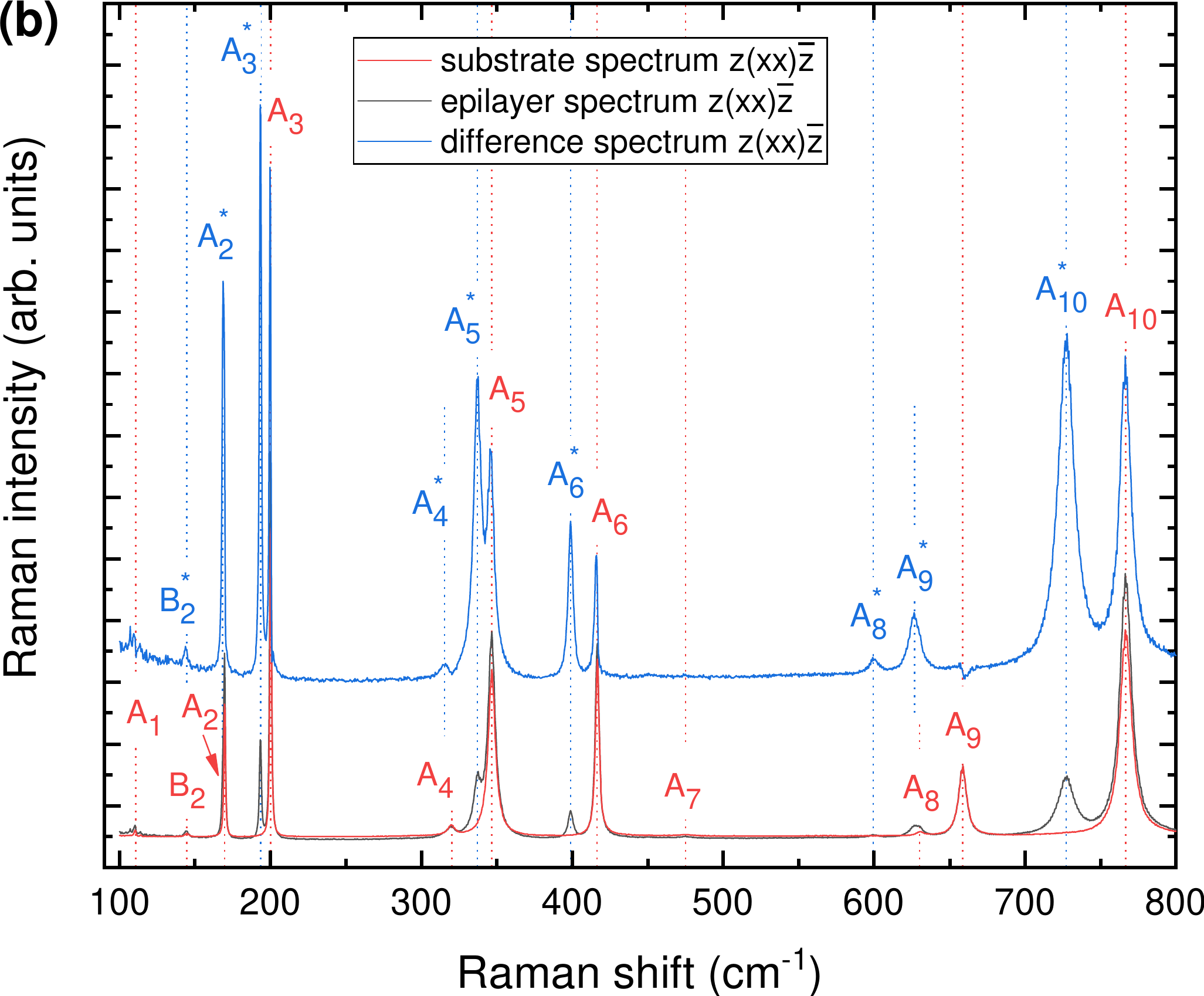}
    \end{minipage}\\
    \vspace{5mm}
    \begin{minipage}{.5\textwidth}
        \flushleft
        \includegraphics[width=81mm]{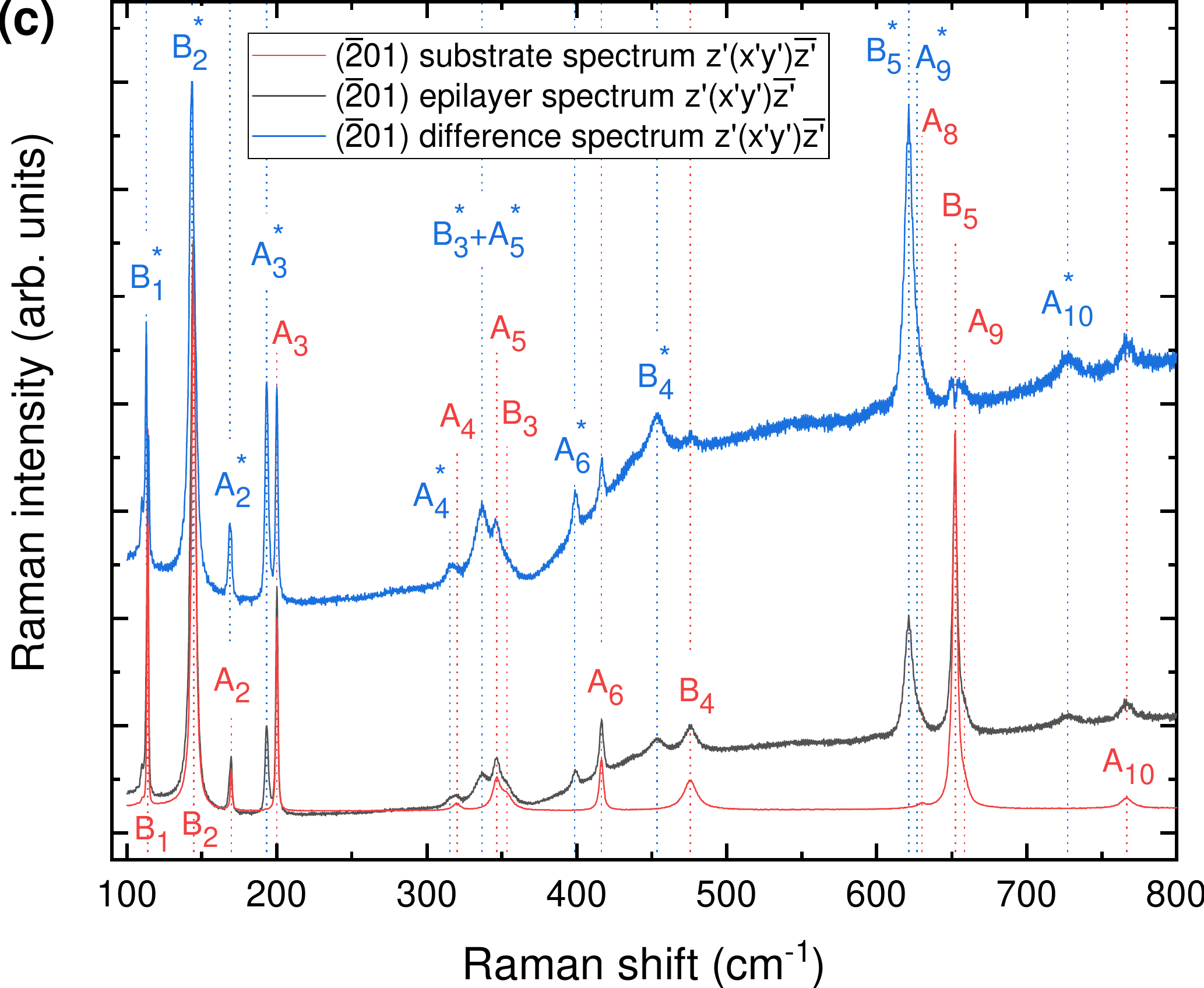}
    \end{minipage}%
    \begin{minipage}{0.5\textwidth}
        \flushright
        \includegraphics[width=81mm]{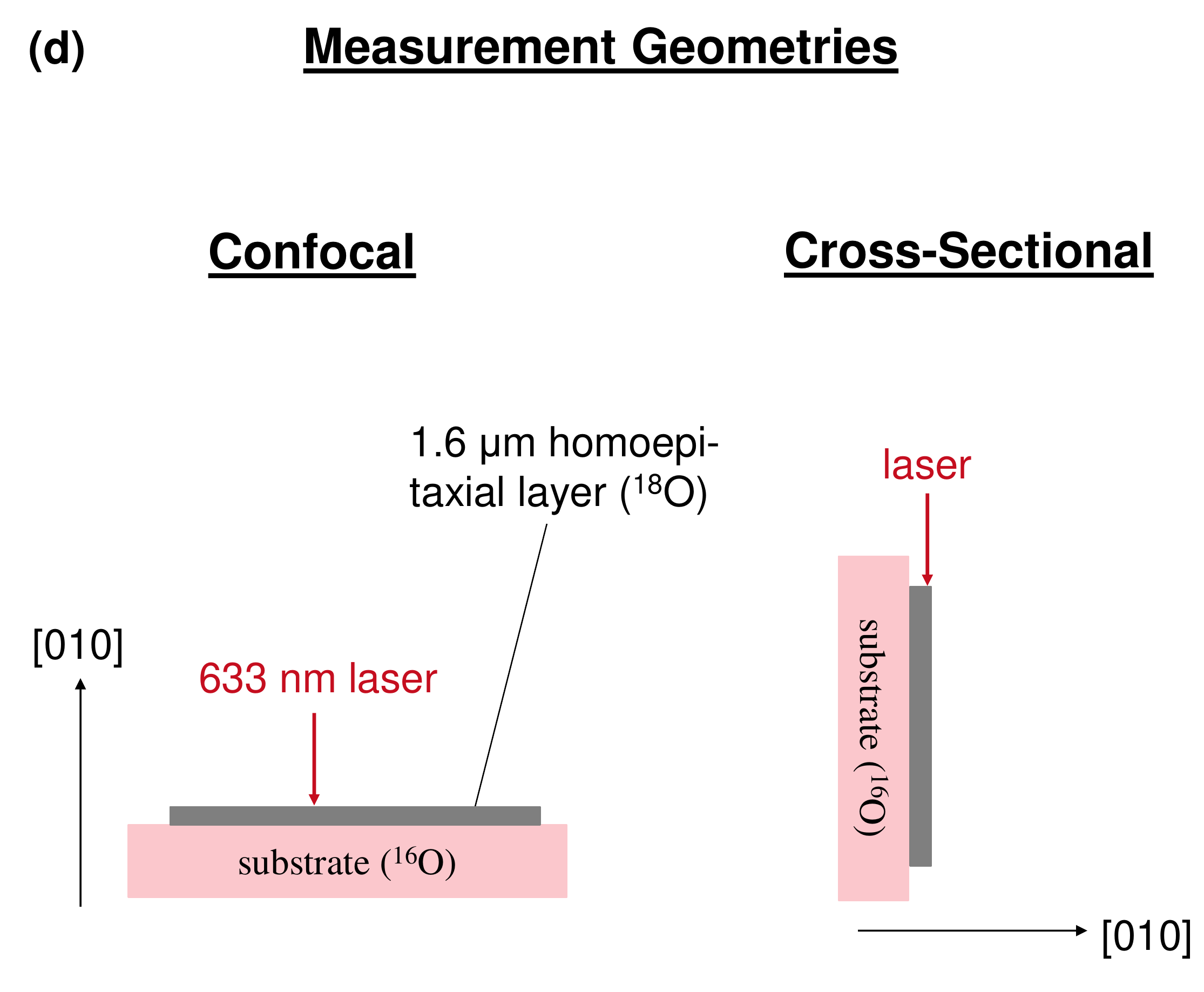}
    \end{minipage}
    \caption{\label{fig:epsart} Raman spectra of the homoepitaxially-grown Ga$_{2}$O$_{3}$ thin film ($^{18}$O) on the (010) plane of a $\beta$-Ga$_{2}$O$_{3}$ substrate with natural ($^{16}$O) isotope abundance.  \textbf{a,b)} Raman spectra in parallel polarization for excitation on the (010)-plane. Substrate spectra (red) were subtracted from the epilayer spectra (grey) to obtain difference spectra (blue), in which $^{18}$O Raman modes (denoted by *) dominate. For clarity, spectra are vertically offset and the difference spectra multiplied by a constant factor. The Cartesian coordinate system $x,y,z$ is chosen such that $z$-axis aligns with the [010] direction, while $y\parallel$[100]. \textbf{c)} Raman spectra of the ($\bar{2}01$)-plane in crossed polarization. The $z'$-axis is parallel to the surface normal, with $x'\parallel$[010] and $y'\parallel$[102]. \textbf{d)} Schematic illustration of sample set-up, plane orientation, and measurement geometry. The (010) plane was accessible in confocal micro-Raman scattering (left). Spectra of the ($\bar{2}$01)-plane were acquired by focusing the laser on the edge of the sample in cross-section geometry (right).  }\label{Figure: Isotope Spectra}
\end{figure}

\section{Results and discussion}

\begin{figure}
\hspace{5mm}
\centering
\includegraphics[width=97mm]{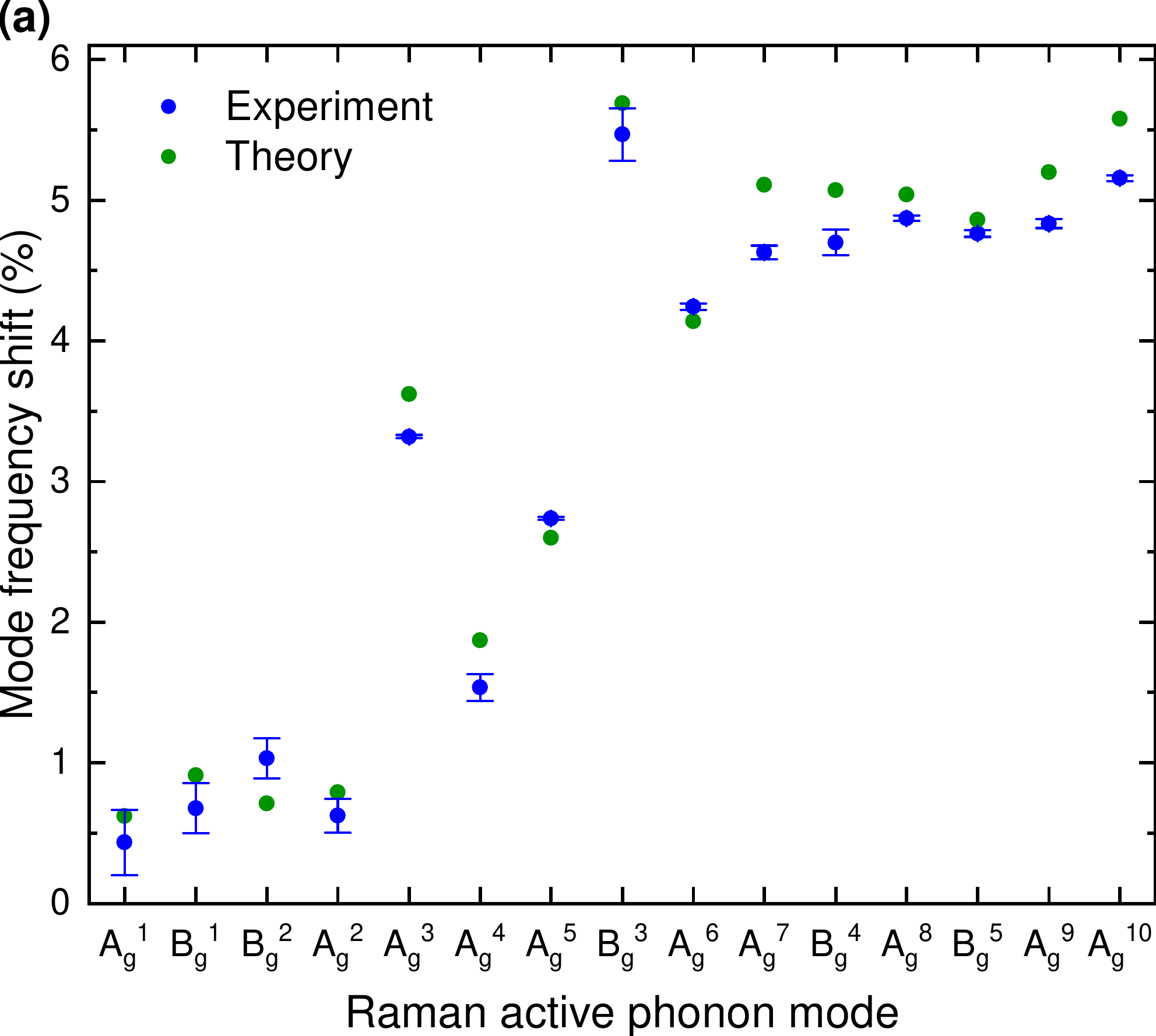}\\
\vspace{5mm}
\includegraphics[width=100mm]{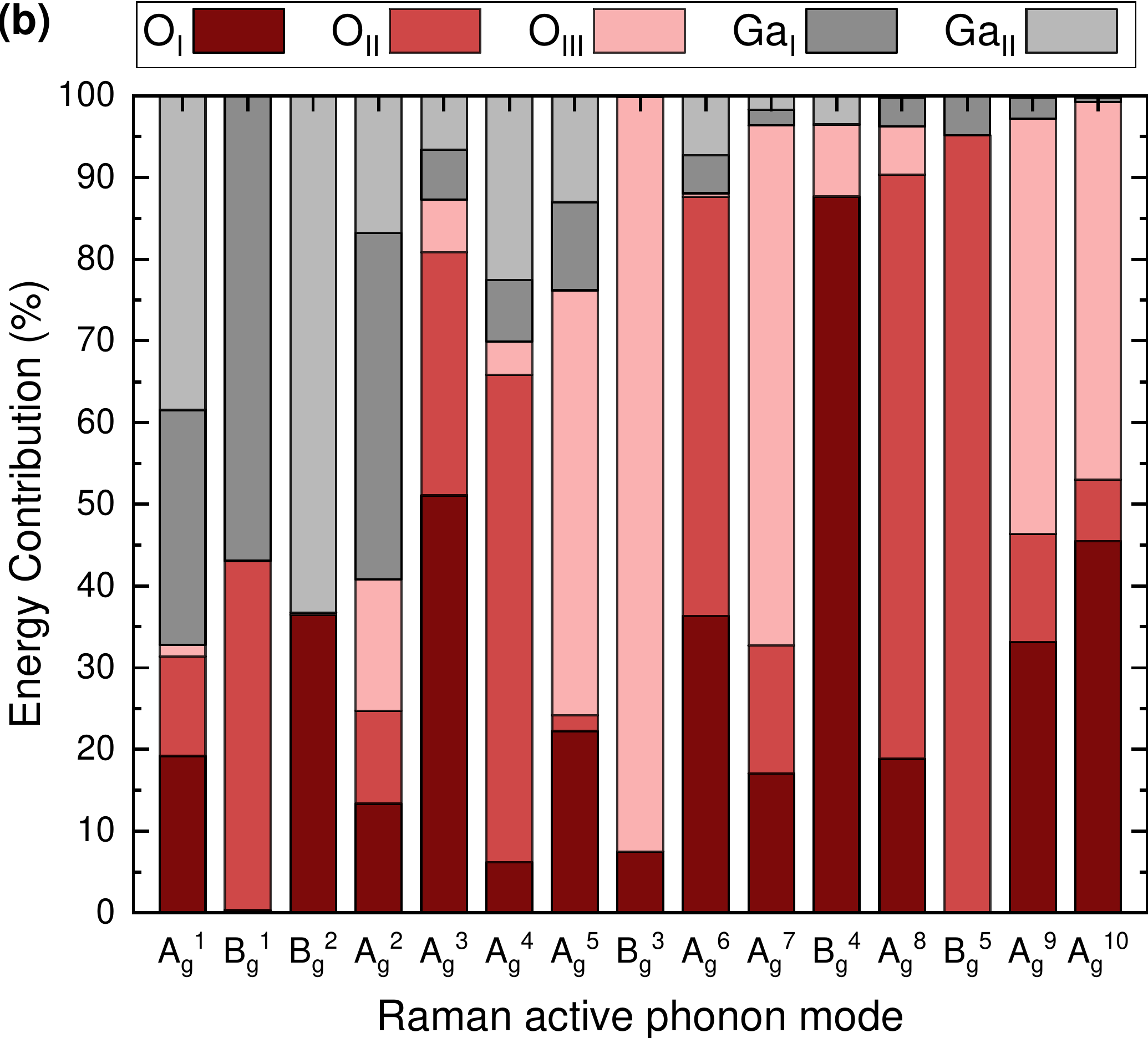}
\caption{(a) Relative mode frequency shifts of Raman modes for $^{18}$O isotope substitution compared to natural isotope distribution based on experimental and theoretical data listed in Table \ref{Table: Peak positions of isotope and natural modes}. (b) Theoretical calculations of relative energy contributions to the total phonon energy corresponding to each O or Ga lattice site. Raman vibrations can be excited by one of the two Ga (grey) or one of the three O (red) lattice sites. The height of a column indicates the relative energy contribution of the corresponding lattice site. }\label{Figure: Relative Mode Frequency Shift}
\end{figure}

\begin{figure}
\includegraphics[width=90mm]{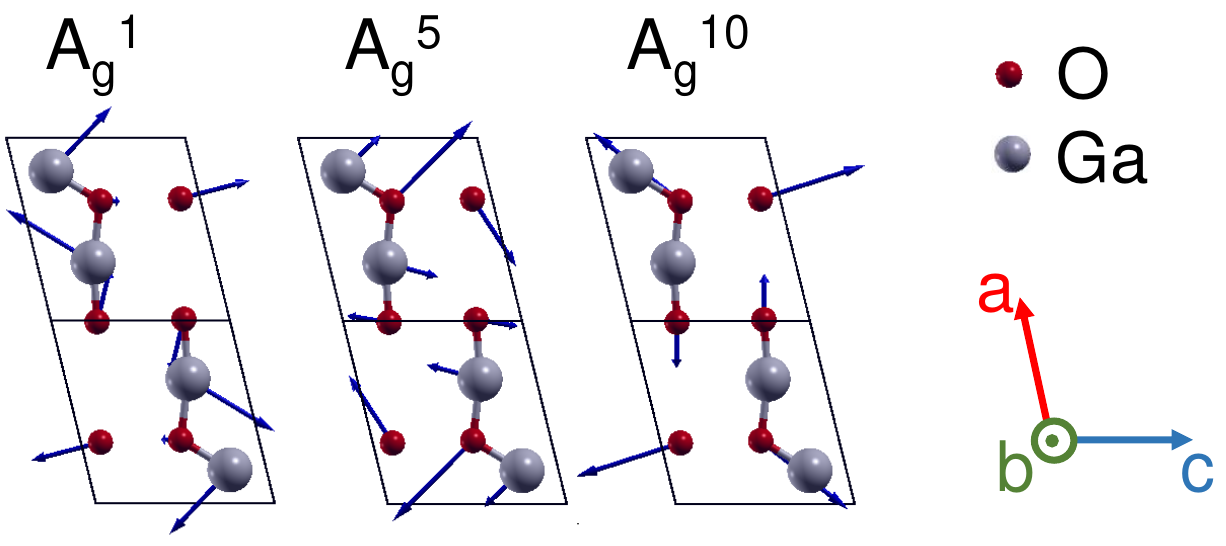}
\centering
\caption{Scheme of the Raman-active $A_{1}$, $A_{5}$ and $A_{10}$ modes within the primitive unit cell of $\beta$-Ga$_{2}$O$_{3}$. Modes are shown in projection on the (010)/$b$-plane. Arrows indicate the displacements of basis atoms, with lengths denoting the amplitude of vibration.}\label{Figure: Scheme of A-modes}
\end{figure}

The Raman-active phonon modes of $\beta-$Ga$_{2}$O$_{3}$ in the $^{16}$O and $^{18}$O isotoplogues (substrate and film, respectively) were acquired in polarized micro-Raman measurements, the results of which are depicted in Fig. \ref{Figure: Isotope Spectra} a-c.
From ToF-SIMS experiments, the dominating presence of $^{18}$O in the homoepitaxial film has been verified: the isotope fraction is 96.3\% and remains constant up to a depth of 1.5 $\mu$m from the surface (Fig. S2 in Supplementary Material). At the interface between film and substrate, the isotope fraction drops over an intermediate region of 300 nm to 0.3\%, which is slightly higher than the natural isotope abundance of 0.2\%. The difference might be caused by diffusion during the deposition process at elevated temperatures. The depth with 50\% of the initial isotope fraction marks the interface of the film resulting in a film thickness of 1.65 $\mu$m (AFM micrography of the layer reported in Fig. S1 of Supplementary Material). 
\newline
The MEXCAT deposition process\cite{Mazzolini,Mazzolini2019,MazzoliniandBierwagen} allows to widen the deposition window of Ga$_{2}$O$_{3}$ to otherwise forbidden growth regimes (e.g. the high T$_{\mathrm{g}}$ and metal-to-oxygen flux ratios used in this work), while at the same time allowing for low incorporation of the catalyzing element.
\newline The primitive unit cell of the monoclinic structure consists of 10 atoms: Ga lattice sites occur in tetrahedral (Ga$_{\mathrm{I}}$) or octahedral (Ga$_{\mathrm{II}}$) coordination, with three nonequivalent oxygen sites (O$_{\mathrm{I}}$, O$_{\mathrm{II}}$ and O$_{\mathrm{III}}$) in between. At the $\Gamma$-point, these correspond to 27 optical phonons: \cite{Dohy,Kranert,Onuma}
$$\Gamma^{\mathrm{opt}}=10A_{g}+5B_{g}+4A_{u}+8B_{u},
$$
15 modes are Raman-active (index $g$), with symmetries $A$ and $B$, and 12 modes are infrared (IR) active (index $u$).
\newline All 15 Raman-active phonon modes of symmetries $A_{g}$ and $B_{g}$ were separated by utilization of the measurement geometries illustrated in \mbox{Fig. \ref{Figure: Isotope Spectra}d.} The sample was irradiated normally in a confocal setup (left) and on the edge (right) to access the (010) or ($\bar{2}01$) plane, respectively. The choice of the (010) and $(\bar{2}01)$ planes is advantageous as it enables the selected detection of Raman modes with $A_{g}$ or $B_{g}$ symmetry and thus facilitates the separation of modes with closely matching phonon frequencies. For the (010) plane, $B_{g}$-modes are forbidden as determined by Raman selection rules. For excitation normal to the ($\bar{2}01$)-plane, $A_{g}$ modes have vanishing intensity for crossed polarization of incident and scattered light when the polarization of the incident radiation aligns with the [010] direction.\cite{Dohy,Kranert} 
\newline Fig. \ref{Figure: Isotope Spectra}a illustrates the Raman spectra of the (010) plane in parallel polarization between incident and scattered light. Using the Porto notation, the applied scattering geometry can be written as $z(yy)\bar{z}$, where $z$ and $y$ correspond to the [010] and [100] directions, respectively (cf. Fig. \ref{Figure: Unit Cell of beta gallium oxide}).\cite{Kranert,Dohy,Onuma} In this configuration, the $A^{1}_{g},A^{2}_{g},A^{3}_{g},A^{7}_{g}$ and $A^{8}_{g}$ modes have maximum intensity. As we only investigated phonon modes accessible in Raman measurements, the index $g$ denoting a Raman-active mode is omitted in the following discussion. The Raman spectra of the substrate with $^{16}$O (red) and the epilayer with $^{18}$O (grey) were recorded in a confocal set-up by moving the $z$-focus into the substrate (minimum signal contribution 
from the film) and into the film (maximum film contribution).
By subtracting the two spectra from each other, we obtained a difference spectrum (blue), in which the $^{18}$O Raman modes dominate. For clarity spectra are vertically offset and the difference spectrum is scaled by a constant factor. $^{18}$O Raman modes are labeled with * to distinguish them from $^{16}$O modes.
\newline Subsequently, the polarization vectors of both the incident and scattered radiation were rotated 90$^{\circ}$ around the [010] direction. The $x$-axis and the [001] direction confined an angle of 13.77$^{\circ}$ (Fig. \ref{Figure: Unit Cell of beta gallium oxide}), yielding the configuration $z(xx)\bar{z}$, in which the $A_{5},A_{6},A_{9}$ and $A_{10}$ modes are maximal. The same procedures as above were applied to acquire the substrate, epilayer and difference spectra, which are displayed in Fig. \ref{Figure: Isotope Spectra}b.
The objective's large NA (0.9) implies a relaxation of selection rules, as light is also collected from directions other than perpendicular to the (010) plane. Hence, the most-intense $B_{2}$-mode is weakly present in \mbox{Fig. \ref{Figure: Isotope Spectra} b.}
\newline To access the $B$-modes, Raman scattering was performed in a cross-section configuration. The thickness of the film equals approximately twice the diffraction limited extent of the laser of about 800 nm. In order to obtain the Raman spectrum of the homoepitaxial layer, we performed cross-section line scans with 200 nm step size. Based on these linescans, we selected two positions for long integration Raman measurements, one for which the $^{18}$O related Raman modes reach maximum intensity and a second one for which only substrate modes are visible. Subtracting of the resulting spectra yielded the difference spectrum (blue) in \mbox{Fig. \ref{Figure: Isotope Spectra}c.}
We selected an edge, whose surface plane corresponds to the ($\bar{2}$01) plane. For normal incidence $z'$ (cf. Fig. \ref{Figure: Unit Cell of beta gallium oxide}) with crossed polarization in the $z'(x'y')\bar{z}'$ configuration ($x\parallel$[010], $y'\parallel$[102]), Raman selection rules predict vanishing intensity for the $A$ modes and maximum intensity for the $B$ modes. Due to the edge's imperfect preparation, $A$-modes were still present, yet less intense. An intensity ratio of about 3:1 between the most intense $B_{2}$ and $A_{3}$ modes was achieved. All five $B$-modes are hence available in this configuration.  
\newline The spectral positions of $^{16}$O and $^{18}$O Raman modes are derived from the difference spectra in Figs. \ref{Figure: Isotope Spectra} a-c by fitting Lorentzian lineshape functions. The obtained peak positions are listed in \mbox{Table \ref{Table: Peak positions of isotope and natural modes}} for both O isotopes in conjunction with the results of DFPT calculations. A few modes deserve particular attention, as the determination of their spectral positions and relative frequency shifts is challenging due to small frequency shifts or overlapping modes. This applies to the low-frequency $A_{1}$, $B_{1}$, $B_{2}$ and $A_{2}$ modes, for which the frequency shifts are in the order of the respective mode's linewidth.
\newline Furthermore, in the Raman spectra of the substrate (red) shown in Fig. \ref{Figure: Isotope Spectra}c, the $B_{3}$ resides in the right shoulder of the $A_{5}$. As for the $^{18}$O distribution (blue), the two modes are superimposed in a joint widened peak. With the $A_{5}^{*}$ position derived from \mbox{Fig. \ref{Figure: Isotope Spectra}b}, we analyzed this peak by setting the position of the $A_{5}^{*}$ fixed and varying the position of the $B_{3}^{*}$ until the intensity ratio $A_{5}^{*}/B_{3}^{*}$ was equal to the intensity ratio $A_{5}^{}/B_{3}^{}$.
In the $^{16}$O substrate spectrum the $B_{5}$ is closely neighbored by the $A_{9}$. The $B_{5}^{*}$ envelope in the $^{18}$O difference spectrum is composed of three individual modes: In addition to the $B_{5}^{*}$, the $A_{9}^{*}$ as well as the $A_{8}$ mode from the substrate lie in close proximity. The intensity of the $A_{8}$ can be regarded as negligible due to its suppression in the difference spectrum.
\newline Distinguishing the $A_{7}$ and $B_{4}$ modes in both isotope compositions is a formidable task in the literature, as the two modes are located at nearly the same frequency. With the exception of one publication,\cite{Onuma} previous experimental works have usually reported both modes at the same frequency or have only assigned one mode,\cite{Dohy,Machon,Kranert} whereas theoretical works have calculated a frequency difference ranging from 0.3 to 13.4 cm$^{-1}$ between these two modes.\cite{Kranert,Dohy,Machon,Liu} 
\newline 
Using a ($\bar{2}01$)-oriented $\beta$-Ga$_{2}$O$_{3}$ sample as reference, we obtained an intensity ratio $B_{4}/B_{2}>1/28$ between the $B_{4}$ and $B_{2}$ mode in the $z'(x'y')\bar{z}'$ configuration. Using this intensity ratio, we conclude that the $B_{4}$ and $B_{4}^{*}$ will contribute more than 70\% to the total intensity of the occurring peaks at $475.9$ cm$^{-1}$ and $453.6$ cm$^{-1}$. Hence, we assign these peaks to the $B_{4}$ and $B_{4}^{*}$, respectively. Analogously, the peak positions of the $A_{7}$ and $A_{7}^{*}$ are determined from the analysis of the (010) spectra in Fig. \ref{Figure: Isotope Spectra}a with a negligible $B_{4}$ or $B_{4}^{*}$ intensity, respectively. Consequently, we obtain a previously unresolved mode spacing of the $A_{7}$ and $B_{4}$ of 0.6 cm$^{-1}$ and 0.3 cm$^{-1}$ for the $^{16}$O and $^{18}$O isotopologues.      
\newline Following this detailed analysis, we were able to determine the spectral positions of all 15 Raman-active phonon modes in the $^{16}$O and $^{18}$O isotopologues of $\beta$-Ga$_{2}$O$_{3}$ (summarized in Table \ref{Table: Peak positions of isotope and natural modes}).
\newline While a slight change in the oxygen mass does not affect the formation of point defects during the MBE growth, altering the mass of one of the two elements of a binary oxide induces a shift in Raman modes, in which atomic vibrations of the respective element are present. Owing to the larger relative mass difference between e.g. $^{16}$O and $^{18}$O compared to stable Ga isotopes, oxygen isotopes produce an elevated frequency shift and are preferably used to study the variation of phonon frequencies in different isotopologues.\cite{Cardona} 
The observed shift of Raman modes towards lower frequencies upon substitution of $^{16}$O with $^{18}$O corresponds to an increase in the isotopic mass.\cite{Cardona,McCluskey}
\newline Table \ref{Table: Peak positions of isotope and natural modes} further lists the absolute and relative frequency shift for each mode. Based on the data displayed in Table \ref{Table: Peak positions of isotope and natural modes}, \mbox{Fig. \ref{Figure: Relative Mode Frequency Shift}a} depicts the experimentally- (blue) and theoretically-determined (green) relative mode frequency shifts for all 15 Raman active modes. Errors in the experimental data originate from the uncertainties in the determined peak positions as described above. A qualitative agreement between experimental and theoretical data is apparent, with a slight overestimation of the frequency shifts of the majority of modes by the DFPT calculations.
\newline The analysis of the experimental Raman mode shifts in Fig. \ref{Figure: Relative Mode Frequency Shift}a reveals several noteworthy results:
(i) the relative shift strongly varies between the different modes with the smallest and largest shift of 0.43\% for the $A_{1}$ and 5.47\% for the $B_{3}$, respectively;
(ii) low energy phonons between 110 and 170 cm$^{-1}$ exhibit weak frequency shifts below 1.03\%, whereas high frequency phonons with Raman shifts above 350 cm$^{-1}$ experience large relative shifts close to 5\% upon O isotope substitution;
(iii) phonons with wavenumbers between 200 and 350 cm$^{-1}$ show intermediate relative shifts, which do not scale linearly with increasing phonon energy. 
\newline Fig. \ref{Figure: Scheme of A-modes} illustrates a scheme of the Raman-active $A_{1},A_{5}$ and $A_{10}$ modes as representatives of the low-energy phonons, phonons of intermediate energies and high-energy-phonons. Modes of $A$ symmetry oscillate within the (010)-plane, with arrows indicating the amplitude of vibration. A scheme of all Raman-active phonon modes is presented in Fig. S3 in the Supplementary Material.
\newline In order to explain the reasons for the observations (i)-(iii), we calculate the relative energy contribution of the three oxygen (O$_{\mathrm{I}}$, O$_{\mathrm{II}}$, O$_{\mathrm{III}}$) and two gallium (Ga$_{\mathrm{I}}$, Ga$_{\mathrm{II}}$) lattice sites to the total phonon energy for each mode \mbox{(Fig. \ref{Figure: Relative Mode Frequency Shift}b).} The size of the relative frequency shift is dependent upon the contribution of O atoms to the mode intensity. While a large relative frequency shift implies the dominance of O lattice site oscillations within a vibrational mode, a small shift corresponds to the absence of the same.  
The occurrence of a frequency shift in each of the 15 Raman modes suggests that O vibrations contribute to each phonon mode, which is confirmed by the fact that O lattice sites are present in each mode's energy contribution. Hence, pure Ga-Ga vibrations do not occur.

\definecolor{Gray}{gray}{0.9}
\begin{table*}
\caption[Spectral positions of Raman peaks of the phonon modes of $\beta-\mathrm{Ga}_{2}\mathrm{O}_{3}$ in natural $^{16}$O and $^{18}$O isotope distribution]{Spectral positions of Raman peaks of the phonon modes of $\beta-\mathrm{Ga}_{2}\mathrm{O}_{3}$ in natural $^{16}$O and $^{18}$O isotope distribution, given in cm$^{-1}$. Peak positions were determined from Figs. \ref{Figure: Isotope Spectra} a-c by fitting Lorentzian lineshape functions. The absolute and relative frequency shifts of each Raman mode are calculated and are given in cm$^{-1}$ or \%, respectively. Relative shifts are rounded to two decimal places, all other quantities on one decimal place. Experimental findings are compared with the results of theoretical DFPT calculations. }
\renewcommand{\arraystretch}{1.3}
\centering
\resizebox{\textwidth}{!}{
\begin{tabular}{|c|c|c|c|c|c|c|c|c|}
\hline
\rowcolor{Gray}\multirow{3}{*}{} & \multicolumn{4}{c|}{Experiment} & \multicolumn{4}{c|}{Theory} \\ \cline{2-9} \rowcolor{Gray}  Phonon mode &   $^{16}$O  & $^{18}$O   & Absolute shift    &  Relative shift   &  $^{16}$O   &     $^{18}$O&  Absolute shift   &  Relative shift   \\
\rowcolor{Gray}   & (cm$^{-1}$)    & (cm$^{-1}$)    &  (cm$^{-1}$)   &  ($\%$)   &  (cm$^{-1}$)    & (cm$^{-1}$)    &  (cm$^{-1}$)   &  ($\%$)     \\ \hline
                 $A_{g}^{1}$        & 110.7   &   110.2   &   0.5     &   0.43    & 106.4 &   105.7   &   0.7    &   0.62  \\ \hline
           $B_{g}^{1}$        & 114.3   &   113.5   &   0.8    &   0.68    & 107.7 &   106.7   &   1.0    &   0.91  \\ \hline
           $B_{g}^{2}$        & 144.9   &   143.4   &   1.5     &   1.03    & 145.0 &   144.0   &   1.0    &   0.71  \\ \hline
           $A_{g}^{2}$        & 169.8   &   168.7   &   1.1     &   0.62    & 163.1 &   161.8   &   1.3    &   0.79  \\ \hline
           $A_{g}^{3}$        & 200.3   &   193.7   &   6.6     &   3.32    & 190.5 &   183.6   &   6.9    &   3.62  \\ \hline
           $A_{g}^{4}$        & 320.3   &   315.4   &   4.9     &   1.54    & 314.0 &   308.1   &   5.9    &   1.87  \\ \hline
           $A_{g}^{5}$        & 347.0   &   337.5   &   9.5     &   2.74    & 345.0 &   336.0   &   9.0    &   2.60  \\ \hline
           
           $B_{g}^{3}$        & 353.4   &   334.1   &   19.3    &   5.47    & 351.4 &   331.4   &   20.0   &   5.69  \\ \hline
           $A_{g}^{6}$        & 416.7   &   399.0   &   17.7    &   4.24    & 384.2 &   368.3   &   15.9   &   4.14  \\ \hline
           $A_{g}^{7}$        & 475.3   &   453.3   &   22.0    &   4.63    & 458.9 &   435.5   &   23.4   &   5.11  \\ \hline
           $B_{g}^{4}$        & 475.9   &   453.6   &   22.3    &   4.70    & 473.3 &   449.3   &   24.0   &   5.07  \\ \hline
           $A_{g}^{8}$        & 630.4   &   599.6   &   30.8    &   4.87    & 620.3 &   589.1   &   31.2   &   5.04  \\ \hline
           $B_{g}^{5}$        & 652.4   &   621.3   &   31.1    &   4.76    & 644.4 &   613.1   &   31.3   &   4.86  \\ \hline
           $A_{g}^{9}$        & 659.0   &   627.2   &   31.8    &   4.83    & 648.5 &   614.8   &   33.7   &   5.20  \\ \hline
           $A_{g}^{10}$       & 767.3   &   727.7   &   39.6    &   5.16    & 751.5 &   709.6   &   41.9   &   5.58  \\ \hline
\end{tabular}}
\label{Table: Peak positions of isotope and natural modes}
\end{table*}

The smallest relative frequency shifts seen in the low-frequency $A_{1}$, $B_{1}$, $B_{2}$ and $A_{2}$ modes correspond to a comparatively low energy contribution from O lattice sites to the respective modes.
An increased frequency shift relative to the low-frequency modes is distinctive of the $A_{3}$, $A_{4}$ and $A_{5}$ modes.
O lattice sites are expected to contribute 76$\%$ to the energy of the $A_{5}$, which is reflected in an observed relative frequency shift of $2.74\%$.
Conversely, a lower frequency shift of 1.54$\%$ for the $A_{4}$ stems from a less prominent energy contribution from O lattice sites ($70\%$).
Oxygen lattice site vibrations are predominant in the $B_{3}$ and the remaining high-frequency modes from $A_{7}$ onward and vary between $97\%$ and $99\%$. Correspondingly, these modes exhibit the greatest mode frequency shifts.
\newline 
Finally, we address the impact of oxygen isotope substitutions on the three distinct oxygen lattice sites. A direct comparison of the observed mode frequency shifts and phonon energy contribution of individual lattice sites indicates that the size of the relative frequency shift depends on which of the three O lattice sites contributes to the vibrational mode. 
The experimental data indicate that replacing $^{16}$O with $^{18}$O isotopes on the O$_{\mathrm{II}}$ lattice site induces an increased frequency shift compared to O$_{\mathrm{I}}$ and O$_{\mathrm{III}}$:
As for the $A_{3}$ and $A_{6}$, oxygen vibrations account for 86\% or 87\% of the mode energy. However, the $A_{6}$, characterized by a more intense contribution from O$_{\mathrm{II}}$ lattice sites and a reduced contribution from O$_{\mathrm{I}}$ sites, has an increased relative frequency shift. 
Equally, oxygen vibrations equal between 96$\%$ and 97$\%$ in the $A_{7}$, $B_{4}$, $A_{8}$ and $A_{9}$ modes. Yet, the observed frequency shift is most profound for the $A_{8}$, where O$_{\mathrm{II}}$ lattice sites prevail. As for the $B_{5}$, the O energy contribution is comprised entirely of O$_{\mathrm{II}}$ sites. Despite a lower overall oxygen contribution (95\%) compared to the $A_{7}$ and $B_{4}$ modes, the $B_{5}$ is characterized by an increased relative frequency shift.

\section{Summary and Conclusion}

In summary, we examined a homoepitaxial $\beta$-Ga$_{2}$O$_{3}$ thin film in the $^{18}$O isotope composition, deposited on top of a substrate with $^{16}$O isotope distribution. Polarized micro-Raman spectroscopy was carried out to record Raman spectra of both isotopologues. By probing the (010) and ($\bar{2}01$) planes, we were able to separate all 15 Raman active phonon modes and determine their phonon frequencies for both isotopologues. The replacement of $^{16}$O with the heavier $^{18}$O atoms resulted in a reduction of the phonon frequency.
We provided the spectral positions of $^{18}$O Raman modes and quantified the absolute as well as relative frequency shifts compared to $^{16}$O Raman modes.
Based on the comparison of experimental data with DFPT calculations of mode frequency shifts and calculation of the relative energy contributions of each lattice site to the total phonon energy, we identified and quantified the atomistic vibrations for each of the five distinct lattice sites which give rise to the 15 different Raman-active phonon modes in $\beta$-Ga$_{2}$O$_{3}$.
A large relative frequency shift seen in the modes of higher frequencies is emblematic of modes governed predominantly by O lattice site vibrations. In contrast, the low-frequency $A_{1}$, $B_{1}$, $B_{2}$ and $A_{2}$ modes exhibit the smallest overall shifts and hence are governed by the vibration of both Ga and O atoms. By determining dedicated Raman modes dominated by one of the three inequivalent O-sites and relating their relative frequency shifts to the calculated relative amount each lattice site contributes to a respective mode's energy, we conclude that substituting $^{16}$O with $^{18}$O isotopes on the O$_{\mathrm{II}}$ lattice site results in an elevated mode frequency shift compared to O$_{\mathrm{I}}$ and O$_{\mathrm{III}}$ lattice sites.  
This discovery paves the way for the identification of O-site-related defects by Raman spectroscopy in future studies.
Furthermore, micro-Raman spectroscopy may be carried out to investigate the formation of oxygen vacancies in different lattice sites as a function of different synthesis (e.g. deposition, annealing) conditions.

\section*{Acknowledgement }
The authors thank Maximilian Ries for a fruitful discussion and Thomas Kure for his experimental support.

\newpage
\setcounter{page}{1}
\setcounter{figure}{0}
\setcounter{section}{0}

\renewcommand{\baselinestretch}{1.33} 
\renewcommand{\thepage}{S\arabic{page}}
\renewcommand{\thetable}{S\arabic{table}}
\renewcommand{\thefigure}{S\arabic{figure}}
\renewcommand{\theequation}{S\arabic{equation}}
\renewcommand{\thesection}{S\arabic{section}}

\section{\Large{Supplementary Material}}

The supplementary contains AFM images of the investigated (010)-oriented $\beta$-Ga$_{2}$O$_{3}$ sample, data of the ToF-SIMS measurements used to study the atomic composition of the sample, as well as a scheme of the 15 Raman-active vibrations within the primitive unit cell of monoclinic $\beta-\mathrm{Ga}_{2}\mathrm{O}_{3}$.

\subsection{\label{sec:level1}AFM micrographs}

\begin{figure}[ht]
\centering
\includegraphics[width=150mm]{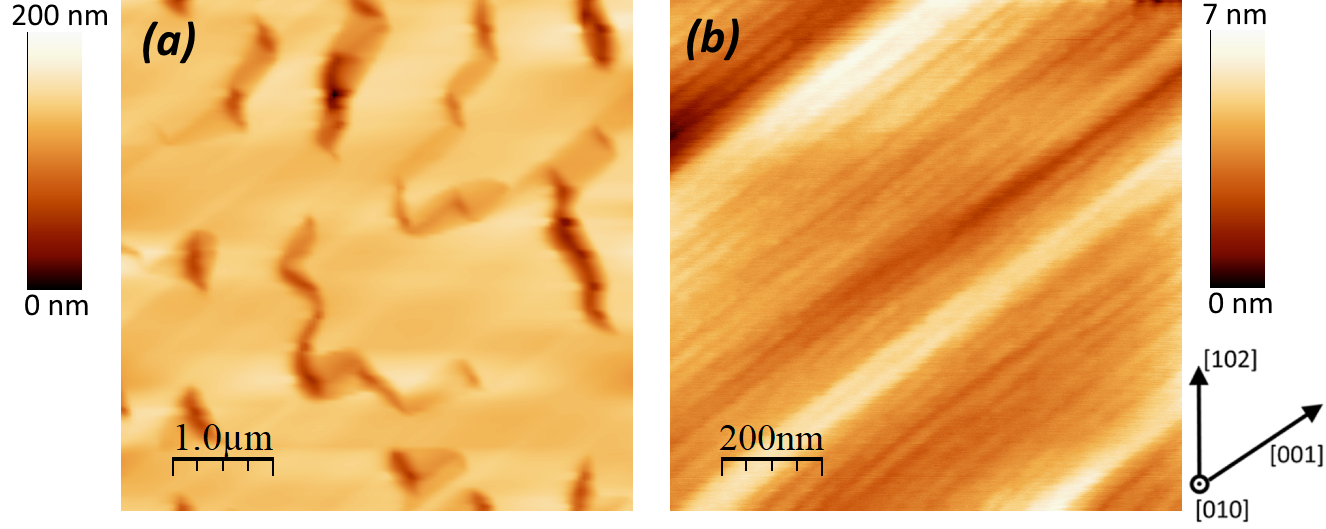}
\caption{\label{Figure: AFM micrograph} $\boldsymbol{(\mathrm{a})}$ 5x5 and $\boldsymbol{(\mathrm{b})}$ 1x1 $\mu$m AFM images of the $1.6$ $\mu$m thick (010) $\beta$-Ga$_{2}$O$_{3}$ homoepitaxial layer deposited with $^{18}$O  isotopes. }
\end{figure}

AFM micrographs evidence a morphology characterized by (i) the presence of deep trenches, almost orthogonal to the [001] in-plane direction, related to island-coalesence growth mechanism (Fig. \ref{Figure: AFM micrograph}a),\cite{MazzoliniandBierwagen} and (ii) (110) facets visible as elongated features oriented along the [001] orientation as a result of the metal-rich growth conditions of the layer (Fig. \ref{Figure: AFM micrograph}b). 

\subsection{ToF-SIMS measurements}

\begin{figure}[ht]
\centering
\includegraphics[width=150mm]{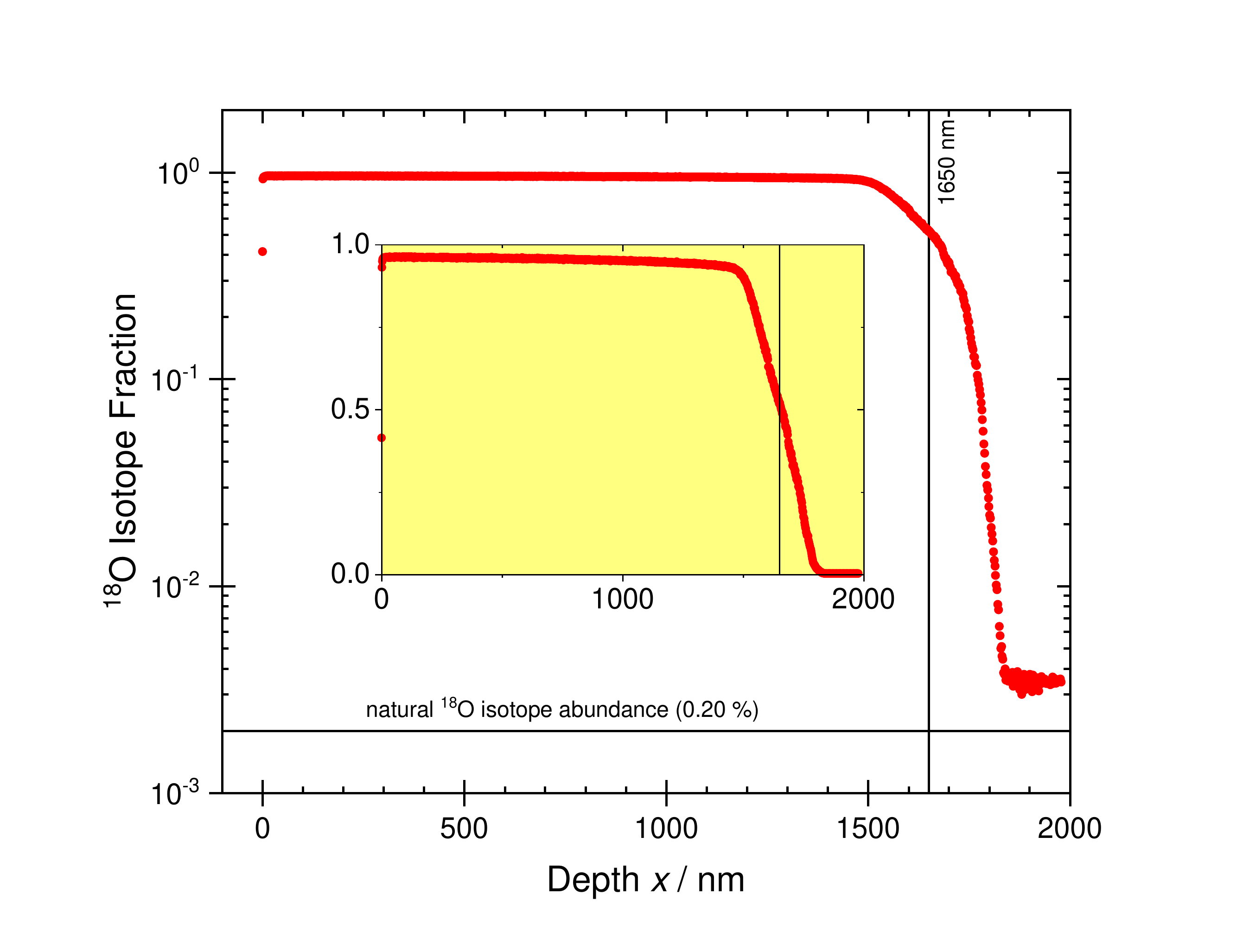}
\caption{\label{Figure: ToF-SIMS} ToF-SIMS analysis of the thin film. $^{18}$O isotope fraction analyzed with 25 keV Ga$^{+}$ analysis beam and 2 kV Cs$^{+}$ sputter beam. The interface (1650 nm) is obtained from the depth of 50\% of the film’s maximum isotope fraction. The inset shows the same graph with linear scale. }
\end{figure}

\mbox{Fig. \ref{Figure: ToF-SIMS}} illustrates the ToF-SIMS depth profiles for the isotope fraction of $^{18}$O. The $^{18}$O isotope fraction $n^{*}$ is calculated by the SIMS intensities: 

\begin{equation}
    n^{*}=\frac{I(^{18}\mathrm{O}^{-})}{I(^{16}\mathrm{O}^{-})+I(^{18}\mathrm{O}^{-})}
   \end{equation}

In the film, 96.3\% $^{18}$O are employed (nominal isotope fraction of the gas: 97.39\%). At the interface, the isotope fraction is decreased over a transient region with an extent of 300 nm to 0.3\%, which is slightly above the natural isotope abundance of 0.2\%.

\newpage
\subsection{\label{Scheme of Raman-active phonon modes}Scheme of Raman-active phonon modes}

\begin{figure}[ht]
\centering
\includegraphics[width=150mm]{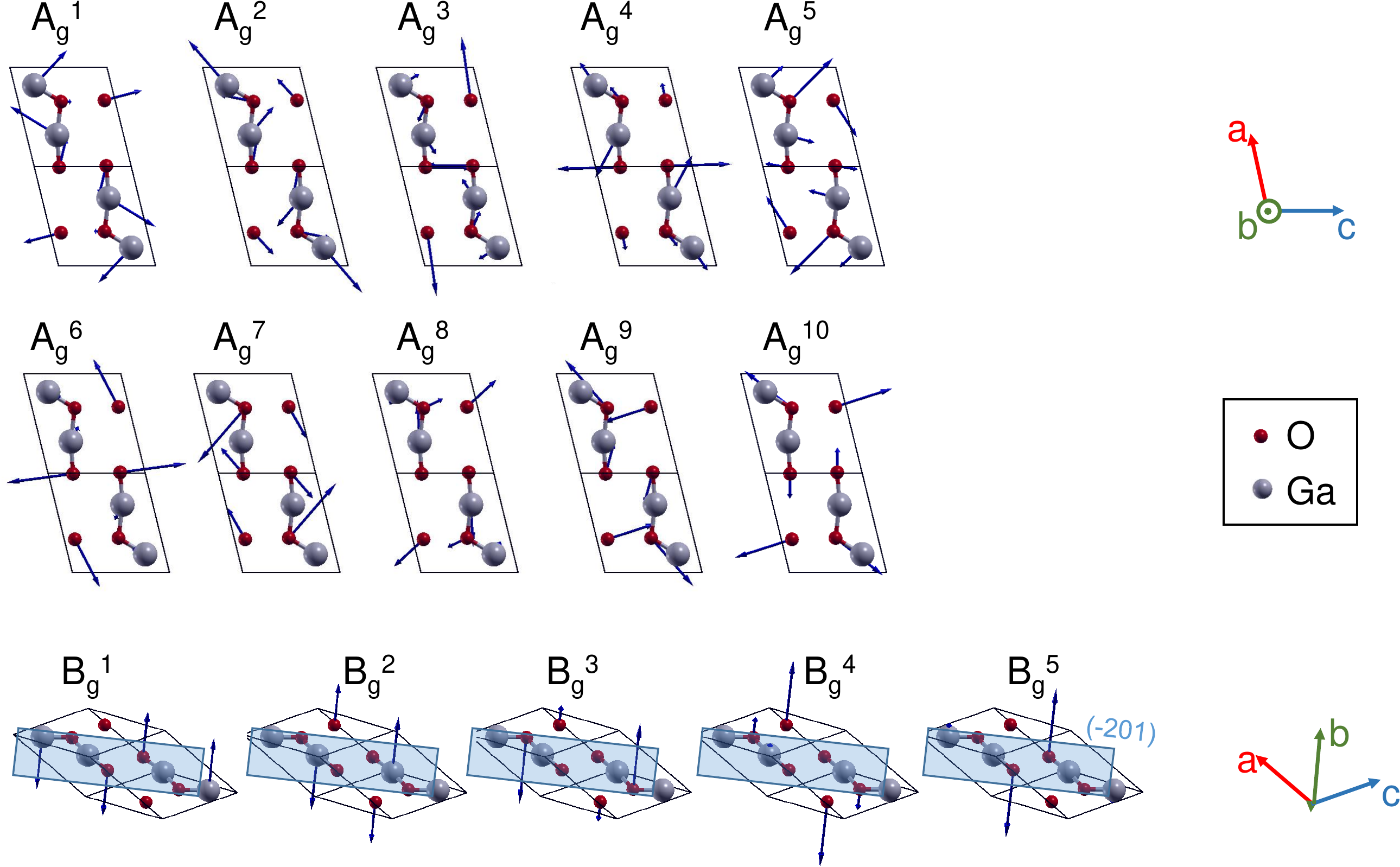}
\caption{\label{Figure: Scheme of Raman active vibrations} Scheme of Raman-active modes within the primitive unit cell of $\beta$-Ga$_{2}$O$_{3}$. $A_{g}$-modes are shown in projection on the $b$-plane. The ($\bar{2}01$) plane (blue) is indicated for the illustration of $B_{g}$-modes. Arrows indicate the displacements of basis atoms, with lengths denoting the amplitude of vibration. }
\end{figure}

For the monoclinic crystal structure of $\beta$-Ga$_{2}$O$_{3}$ there are 15 Raman-active phonon modes (10 with $A_{g}$ and 5 with $B_{g}$ symmetry), the schemes of which are illustrated in Fig. \ref{Figure: Scheme of Raman active vibrations}.
$A_{g}$-modes are presented in projection on the b-plane. Arrows indicate the displacements of the corresponding atoms, with the length of the arrows representing each atom's amplitude of vibration. $A_{g}$-modes are seen to oscillate within the b-plane, whereas modes of $B_{g}$-symmetry vibrate perpendicular to the same. 
 
\clearpage
\newpage
\setcounter{page}{1}
\pagenumbering{roman}
\bibliographystyle{unsrt}  


\end{document}